# A plastic feedthrough suitable for high-voltage DC femtosecond electron diffractometers

Patrick Gicala, Ariel A. Petruk, Nicolás Rivas[#], Sam Netzke, Kostyantyn Pichugin, German Sciaini[*]

*The Ultrafast electron Imaging Laboratory, Department of Chemistry, University of Waterloo, Waterloo Ontario, N2L 3G1, Canada.*

#Present address: Nuclear Engineering Group, McMaster University, Hamilton, Ontario, L8S 4K1, Canada.

*Correspondence:* gsciaini@uwaterloo.ca

Highly energetic ultrashort electron bunches have the potential to reveal the ultrafast structural dynamics in relatively thicker *in-liquid* samples. However, direct current (DC) voltages higher than 100 kV are exponentially difficult to attain as surface and vacuum breakdown become an important problem as the electric field increases. One of the most demanding components in the design of a high-energy electrostatic ultrafast electron source is the high voltage feedthrough (HVFT), which must keep the electron gun from discharging against ground. Electrical discharges can cause irreversible component damage while voltage instabilities render the instrument inoperative. We report the design, manufacturing, and conditioning process for a new HVFT that utilizes ultra-high molecular weight polyethylene (UHMWPE) as the insulating material. Our HVFT is highly customizable, inexpensive, and has proven to be effective in high voltage applications. After a couple of weeks of gas and voltage conditioning, we achieved a maximum voltage of 180 kV with a progressively improved vacuum level of $1.8 \times 10^{-8}$ Torr.

## I. INTRODUCTION

Time-resolved femtosecond electron diffraction has emerged as a powerful technique for monitoring ultrafast structural dynamics. Over the last four decades the time resolution of tabletop direct current (DC) femtosecond electron diffraction (FED) setups has improved by nearly a factor of one thousand with accelerating voltages reaching ~ 100 kV.[1–24] Increasing the kinetic energy of ultrashort electron bursts is highly desirable; not only does it reduce multiple scattering effects in FED studies of crystalline solid samples, but it also enables the investigation of ultrafast structural dynamics in solution phase.

The implementation of radio-frequency mega-electron volt (MeV) ultrafast electron diffraction in combination with sub-micron thick liquid sheet jets have proven to be successful in determining the structure of water and other solvents[25,26], and the dissociation process of aqueous $I_3^-$.[27] However, the elastic scattering cross-section ($\sigma_{el}$) of 1 – 100-MeV electrons in water is comparable to that of 300-keV electrons owing the plateau behaviour experienced by $\sigma_{el}$ at high kinetic energies[28].

In order to allow the study of solution-phase structural dynamics using a laboratory scale FED setup provided the availability of high precision closed-rack DC 300 kV power supplies[29], we decided to embark in the development of a 300 kV FED diffractometer (see Fig. 1(a) below and Fig. 14 in reference 30) and a novel nanofluidic cell system optimized for the delivery of solution-phase samples at high flow rates[31].

The focus of this paper is to report the design and conditioning process of our power high voltage feedthrough (HVFT). The HVFT acts an electrical conductor and a mechanical support to power and position the electron source inside the FED vacuum chamber. Therefore, the HVFT must be vacuum tight and withstand the pressure differential. It is often the case that as the voltage increases, HVFTs experience surface electrical breakdowns that can lead to damage of the HVFT materials and/or the electron gun assembly.

The lack of readily available and reliable HVFTs rated for 300 kV DC for our application, and a negative experience with outsourcing to a third-party HVFT vendor, motivated us to design and fabricate our HVFT in house. It should be mentioned that there are several efforts devoted to the development and implementation of ~ 100 – 500 kV DC photoelectron guns[32–36] for various applications, including nuclear physics experiments with polarized electron beams[37], free electron lasers[38,39] and energy recovery linacs[40]. These electron sources usually operate with high average current and high bunch charge of the order of tens to hundreds of mA and tens to hundreds of pC ($10^8 – 10^9$ electrons per bunch), respectively. These electron sources often employ highly efficient photocathode materials, e.g., negative affinity GaAs-based and alkali-antimonide. Such cathode materials are very sensitive to moisture and require ultrahigh vacuum conditions, i.e., pressures of the order of $10^{-11}$ Torr to achieve a reasonable lifetime[41]. This vacuum level demands the use of ceramic-based high voltage feedthrough insulators, which present negligible degassing and allow chambers to be backed at relatively high temperature[36]. In contrast, the electron sources used in compact DC FED setups deliver ultrashort electron bursts carrying out about $10^3 - 10^5$ electrons per bunch, which are commonly produced via stable metallic (e.g., gold) photocathodes [11,42] that operate well at moderate vacuum, relaxing the environment conditions and enabling the exploration of other HV insulating materials.

HIGH VOLTAGE FEEDTHROUGH DESIGN

A. Material Selection

We initially produced an HVFT prototype made of a 1-meter-long tube of alumina ceramics – a common material used in HVFTs. While alumina is an effective insulator (Table 1), it failed after reaching 160 kV following a large discharge that led to the fracture of the ceramic body, a common failure in ceramic HVFTs. This experience in addition to the challenges associated with the machining of alumina and the high cost of machinable ceramics prompted us to explore a different material for the construction of our HVFT.

Our feedthrough, dubbed Polyethylene Insulator for Electron Sources (PIES) is constructed from ultra-high-molecular-weight polyethylene (UHMWPE); a plastic polymer of the same family as the insulating layer (cross-linked polyethylene) implemented in HV cables[43].

Table 1: General material properties of UHMWPE and Alumina-based ceramics.

| Property | UHMWPE[44,45] | Alumina[46,47] |
|---|---|---|
| Volume Resistivity | $\sim 10^{16}$ Ω cm | $\sim 10^{14}$ Ω cm |
| Relative dielectric Permittivity | $\sim 2$ | $\sim 9$ |
| Dielectric Strength | $\sim 40$ kV/mm | $\sim 15$ kV/mm |

Our review of the pertinent literature revealed a growing body of research exploring the use of UHMWPE in the construction of HVFTs in liquid argon time projection chambers[48–50] and HVDC transmission systems[45,51]. In fact, it was recently demonstrated that UHMWPE worked well at 300 kV in liquid argon[50], which suggests that UHMWPE may be suitable for use in FED instruments if degassing does not become a major issue.

UHMWPE material is also known as high-modulus polyethylene; it has extremely long chains, presents low moisture absorption, and is inexpensive, highly machinable and structurally stable in our vertical configuration. This makes customizing the HVFT straightforward, providing a faster turnaround between possible design modifications.

Fig. 1(a) shows the vertical cross section of our FED apparatus, containing the metallic photocathode attached to HVFT the via a connecting rod. The anode plate was removed to maximize the distances from the photocathode surface to all grounded surfaces during HV conditioning. The photocathode surfaces were manually polished to mirror like finishing (Fig. 1(c) and 1(d)) utilizing a variety of sanding sheets and lapping pastes. All pieces were finally cleaned by sonication for several hours in acetone and isopropanol.

Some concerns about using UHMWPE were also considered. To conduct FED experiments, vacuum pressures below $10^{-6}$ Torr are usually necessary and the lower the pressure the better to achieve higher voltages. Considering that a similar plastic, high density polyethylene, is known to experience outgassing of nonpolar gasses, residual water, and short aliphatic carbon chains[52], the use of UHMWPE posed some concerns. If degassing is found to be severe, experimental results may suffer. We observed, however, that following about a week of exposure to high vacuum conditions the pressure level in our FED system improved and reached 1.8 × $10^{-8}$ Torr, which is approximately the minimum value attainable in our FED chamber with double and single o-ring seals. This observation made UHMWPE a good candidate for the construction of our HVFT. In addition, its low dielectric constant (closer to vacuum) helps to reduce the electric field strength near triple junctions[53].

B. Description of Design and Field Calculations

One of the most common problems we encountered with our previous HVFTs, and electron sources is breakdown along the insulator surface. A common method to circumvent this issue in ceramic bushings used in outdoor power distribution lines and high voltage feedthroughs employed in liquid argon is to increase the creepage distance to prevent flashovers[48]. Although, it is argued that increasing the creepage distance is not a valid design principle in vacuum[54], it may result necessary to reduce the electric field parallel to contaminated surfaces[54] that could lead to deteriorated vacuum via outgassing. We therefore decided to add a repeating groove across the bottom portion of PIES. Note that such convolutions are likely unnecessary since the vacuum pressure achieved was surprisingly lower than expected. We used a UHMWPE tube of ~ 10 cm in diameter to maintain a wall thickness equal or greater than 2 cm for the HVFT. However, due to the dimensions of the HVFT port in the top flange of the FED chamber, a tube 100 mm in diameter could not be accommodated for the full length.

Therefore, we designed PIES into two parts: the top half is 66 mm in diameter, and the bottom half is ~ 102 mm in diameter. These two parts are joined together by a thread machined directly into PIES, which is sealed using an o-ring. PIES is assembled first by attaching the top half to the top flange and inserting it into the chamber, and then the bottom half is threaded into the upper half. Both halves of PIES were bored with a 22 mm hole to accommodate the HV cable. Moreover, the selected overall length of PIES (~ 1.3 m), was required for positioning the electron source in a particular location inside of the FED chamber. A shorter HVFT (e.g., the bottom half) is anticipated to confer similar HV insulation performance.

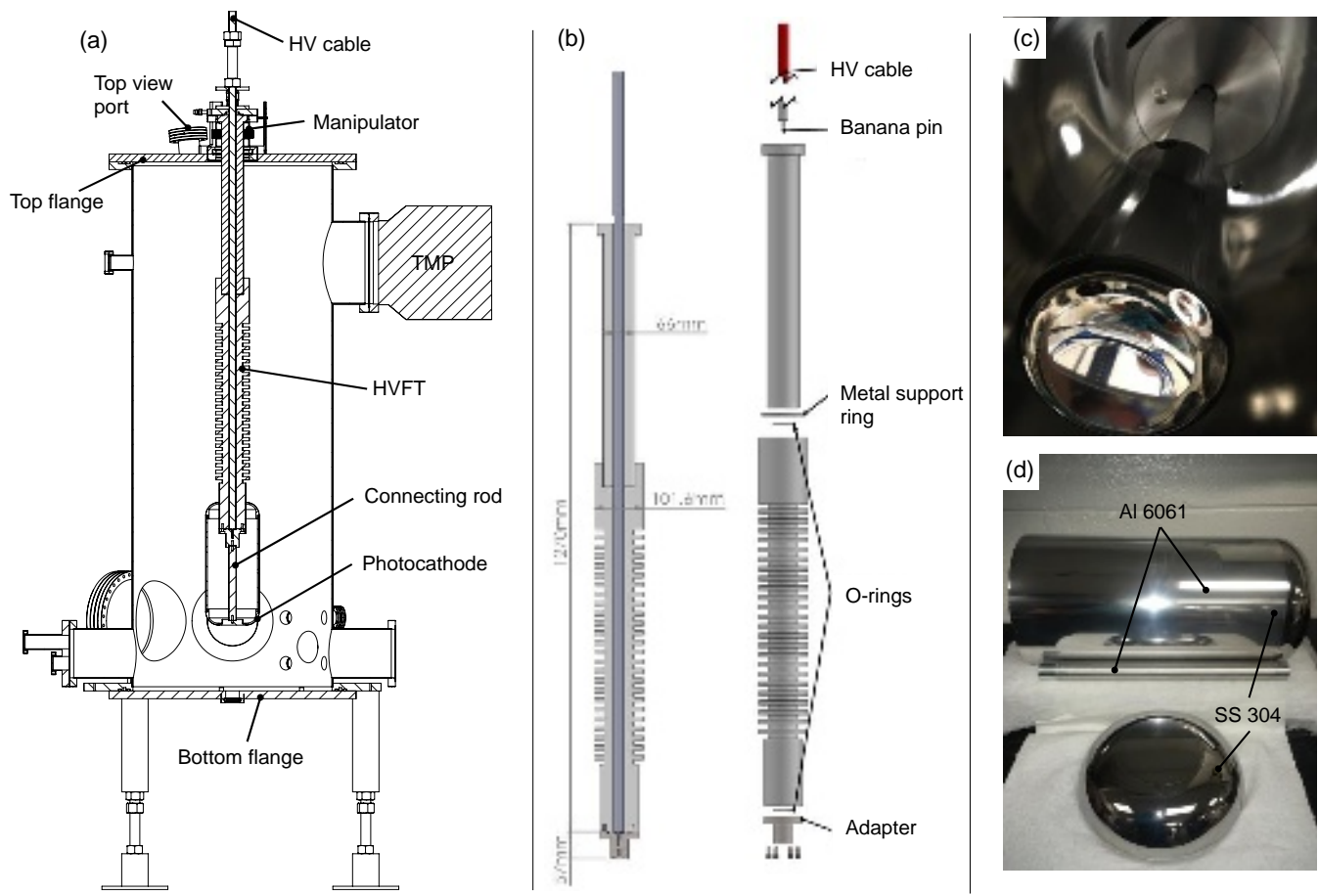

FIG. 1. (a) Vertical cross section of our FED chamber including the main components used during gas and HV conditioning. 3D-computer aided designs (CADs) were generated in SOLIDWORKS[55]. TMP = Turbo molecular pump. (b) 3D-CAD of PIES assembly illustrating the main components of our HVFT. The top and bottom halves of PIES are threaded and sealed using an o-ring. The HV cable plugs directly into the stainless-steel adapter, which is attached using six M6 PEEK screws and sealed using an o-ring. The metal support ring rests between the underside of the top plastic rim and mounting flange, shown in panel (a). (c) Photograph of PIES assembly taken after removing the bottom flange. (d) Photograph of the photocathode component with connecting rod. Al 6061 = Aluminum 6061 and SS 304 = stainless steel 304. Note that all critical photocathode surfaces experiencing high electric fields were machined in SS 304.

The end of the HV cable was modified by soldering a banana pin to the internal copper conductor (Fig. 1(b)). This allows us to easily plug the cable into the coupling adapter. The coupling adapter receives the banana pin and an o-ring to provide sealing upon compression by six M6 PEEK screws which were modified to allow venting, see Figs. 1(a) and (b).

Electrostatic field calculations for our design geometry and selected materials were performed in COMSOL Multiphysics[56]. The electric field map obtained with our photocathode head at a potential of -300 kV with respect to ground is shown in Fig. 2. The maximum surface electric was found to be 5.5 MV/m, which is below the empirical value of ~ 10 MV/m for vacuum breakdown induced by field emission from conditioned surfaces, which has been found by several groups[33–36,42].

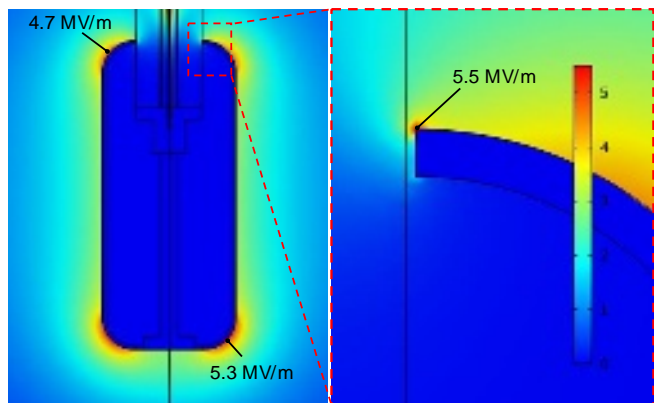

FIG. 2. Electrostatic field calculations performed with COMSOL Multiphysics[56]. The maximum calculated gradient is 5.5 MV/m with the photocathode at -300 kV and it is found at the edge of the metallic photocathode near the insulator-vacuum interface. Note in left panel that there is a small gap of 2 mm between the conductor and the UHMWPE HVFT surface. This gap allows venting of the photocathode interior during pumping down.

### C. Manufacturing

PIES was manufactured at the University of Waterloo by our expert technical services group. They were able to bore a continuous hole through both halves of PIES that matched with a high degree of precision. The surface of the feedthrough was machined to a smooth finish. High vacuum is a necessary condition for FED; therefore, it is imperative that PIES maintains a proper seal to prevent leakage into the chamber. The gap between the HV cable and the inner wall of the HVFT was filled with HV silicone oil, courtesy of CSL Silicones Inc.[57].

After passing a Helium leak test (Fig. 3), PIES was assembled into the FED chamber and a vacuum level of ~ $5 \times 10^{-7}$ Torr was reached within a couple of hours with our 2200 L/s turbomolecular pump.

## II. CONDITIONING OF PIES

Regardless of how well optimized the design of HVFTs is, discharges and instability will occur during their initial use (Fig. 3). Breakdowns can occur between the cathode and grounded components, as well as across the surface of the HVFT insulator. In order to increase the breakdown voltage, HVFTs must be conditioned[58,59]. Conditioning often helps eliminate field emitters, smoothing the surfaces and enabling higher electric fields[60]. We employed two well-known forms of conditioning known as electrical stress and gas conditioning. A 300 kV negative polarity Heinzinger PNChp DC high voltage power supply was implemented to energize the system. The HV cable is a Heinzinger HVC200, rated for use up to 300 kV DC.

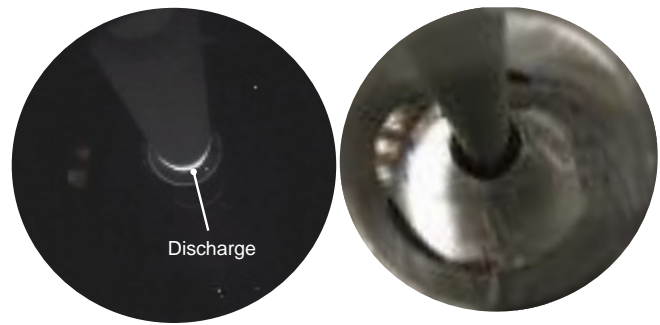

FIG. 3. (a) Photograph of PIES with photocathode inside of the chamber during conditioning. Voltage breakdowns seem to initiate at the tripe-point-like edge of the photocathode (maximum field gradient shown in Fig. 2). This type of discharging was observed with relative high frequency at voltages near 180 kV after gas and HV conditioning. (b) Photograph taken for reference with the bottom flange open.

### A. Gas Conditioning

PIES was initially conditioned utilizing helium at low pressures. Gas conditioning with noble gases is thought to improve the breakdown voltage via ion implantation. This process is described thoroughly in the literature.[61,62] In short, it is believed that by ionizing helium the charged particles accelerate into the cathode and assist in the degassing process as well as the blunting of field emitters. The pressure of gas inside the chamber is difficult to determine, as the gauge is not accurate in measuring pressures above $10^{-1}$ Torr and is sensitive to the residual gas used. In any case, the gauge read out 2 Torr and $10^{-2}$ Torr for phases 1 and 2 respectively. The power supply and He-gas pressure parameters used for the Phase I & II gas conditioning are summarized in Table 2.

Table 2: Parameters for helium gas conditioning of PIES.

| Parameter | Phase I | Phase II |
| --- | --- | --- |
| Gas Pressure | 2 Torr | $10^{-2}$ Torr |
| Current | 200 µA | 200 µA |
| Voltage | 5 kV | 5 kV |
| Time | 24 hrs | 48 hrs |

### B. Electrical Stress Conditioning

Electrical stress conditioning is the most common method of conditioning. It consists of repeatedly increasing the voltage as discharges take place. We monitor discharges by observing the voltage and current fluctuations in the HV power supply as well as changes in vacuum level of the FED chamber. Our stress conditioning protocol dictates that if discharges occur, the voltage is lowered until stable conditions are observed and left there for approximately 10 minutes. The voltage is then steadily increased just over the voltage where last evident discharge took place and left there for a moment. If stable conditions are observed after approximately 10 minutes, conditioning continues by increasing the voltage further. It is thought that this action blunts field emitters on the cathode surface. After one week of conditioning using this method, a maximum voltage of 180 kV was reached. We would like to mention that further conditioning is expected to continue improving the high voltage rating of PIES and our electron source. However, we decided that developing an intermediate acceleration stage (biased at -150 kV) will be the most suitable way to obtain 300 keV electrons while maintaining low radiation levels and attaining the long-term stability required to perform time demanding FED experiments.

### III. CONCLUSION AND FUTURE OUTLOOK

We have successfully designed and manufactured a high voltage feedthrough for our FED instrument that has been conditioned up to a voltage of 180 kV. The biggest advantages of UHMWPE are the ease of machining that provides the opportunity to explore different HVFT design geometries and its low electrical permittivity[53]. UHMWPE proved to be a good choice for the development of HVFTs for our application as it behaves well under vacuum conditions and degassing does not appear to be an issue. Moreover, UHMWPE presents excellent insulating and good mechanical properties which can be modified by the addition of different filling materials to gain control on surface resistivity and hardness values, opening the doors to the implementation of hybrid designs, i.e., combinations of different insulating materials. PIES operates well at voltages below 180 kV and has shown no visible damage. Future plans involve the improvement of the coupling adapter by adding a triple-junction screen[36] or rounding of the top edge of the photocathode head, and the inclusion of the intermediate stage to enable stable long-term operations with 300 keV electrons. We believe that our FED community will greatly benefit from the use of custom made HVFT based on UHMWPE, which has shown to be an excellent choice for most common 100-kV FED setups.

### IV. ACKNOWLEDGEMENTS

We acknowledge the support provided by the National Science and Engineering Research Council of Canada, the Canada Foundation for Innovation, and the Ontario Research Foundation. G.S. is grateful for the support provided by the Canada Research Chairs program. We would like to thank Prof. Tong Leung (Watlab) for lending us the Helium leak detector and pump trucks.

### V. DATA AVAILABILITY

Drawings are available from the corresponding author upon reasonable request.

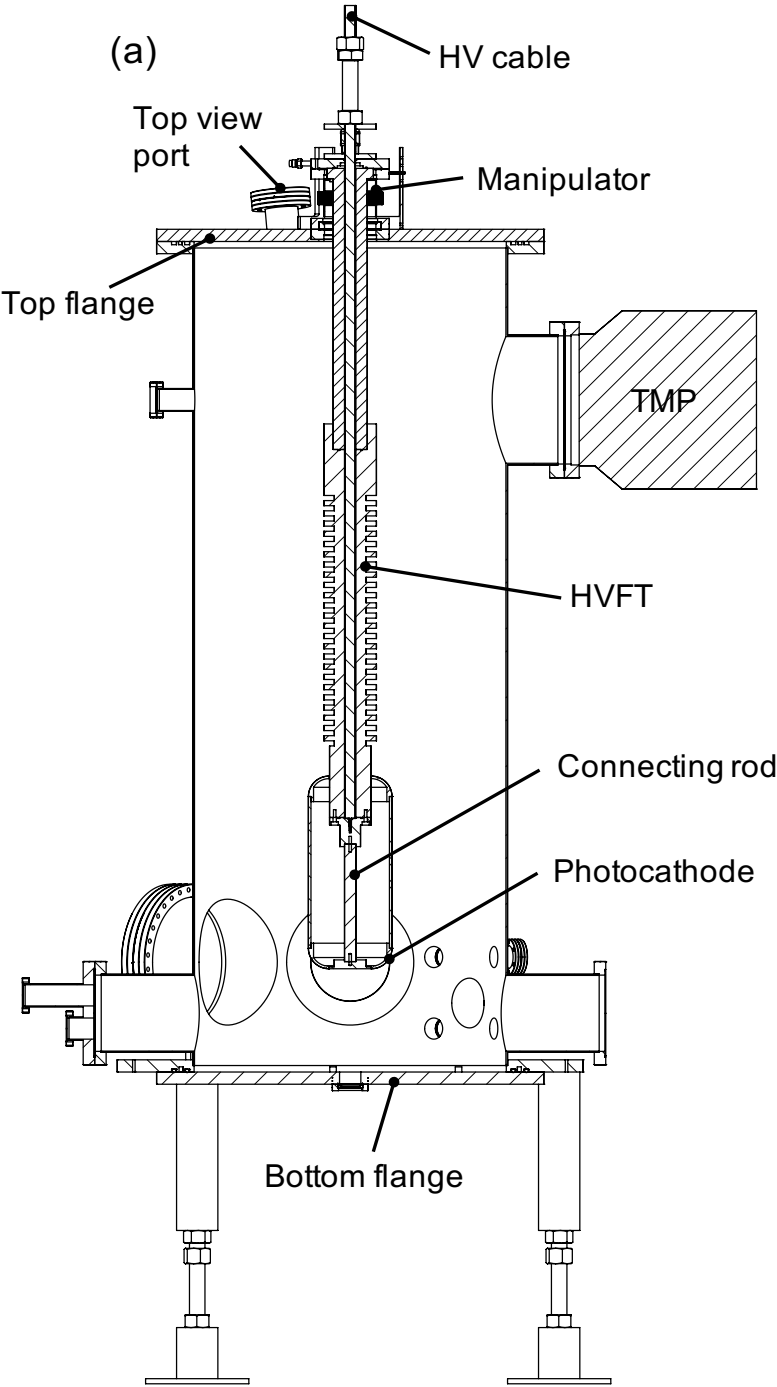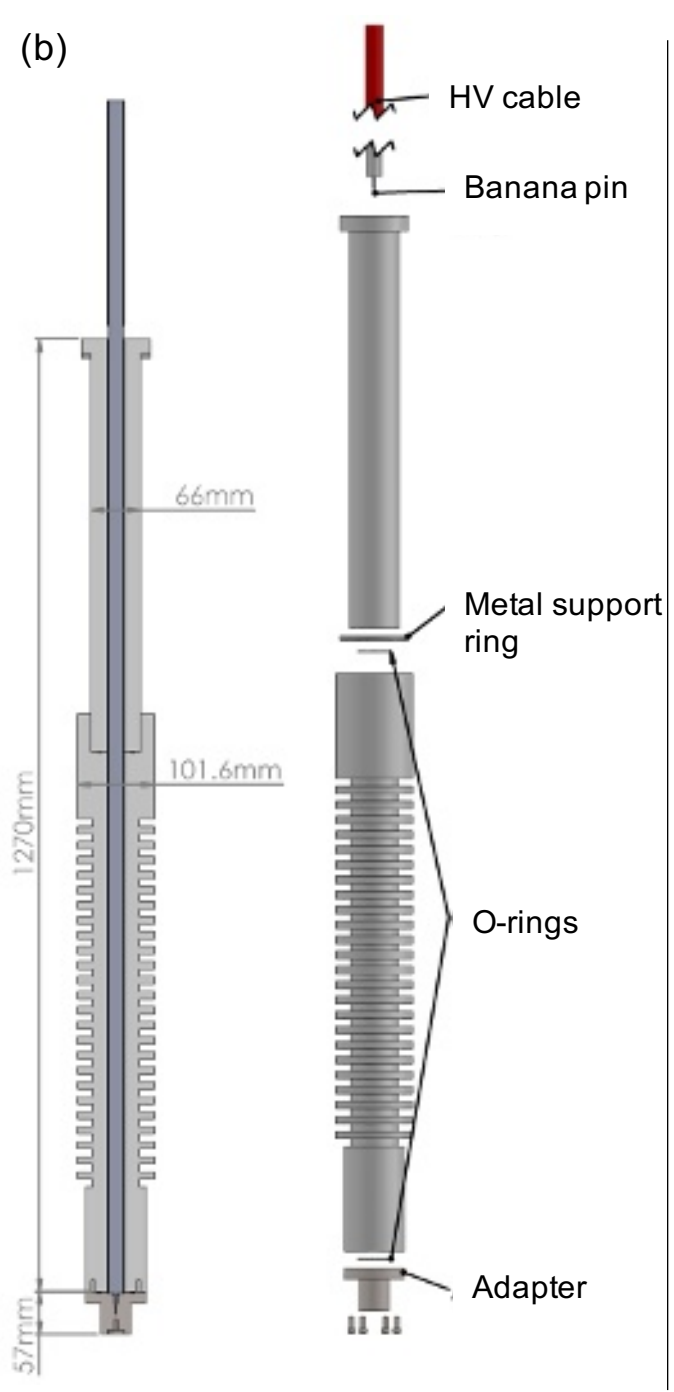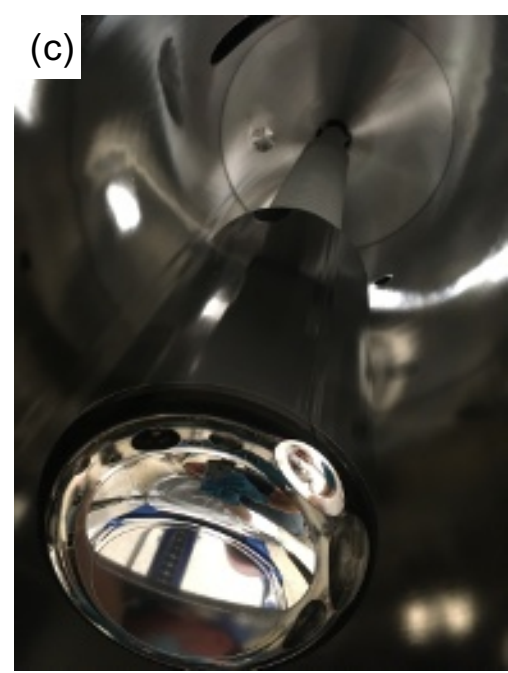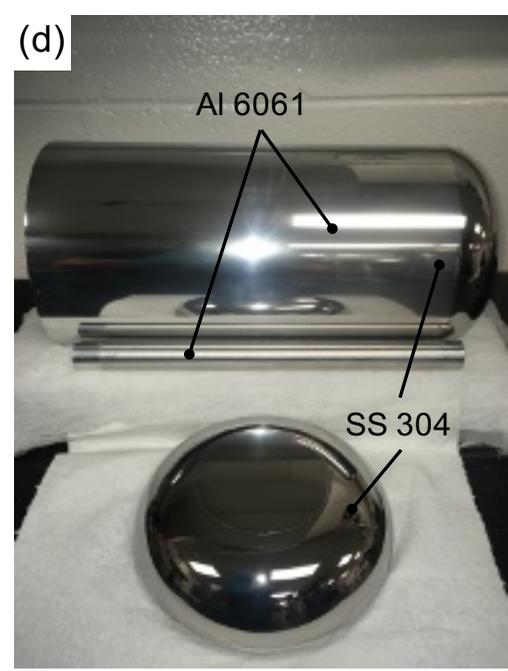

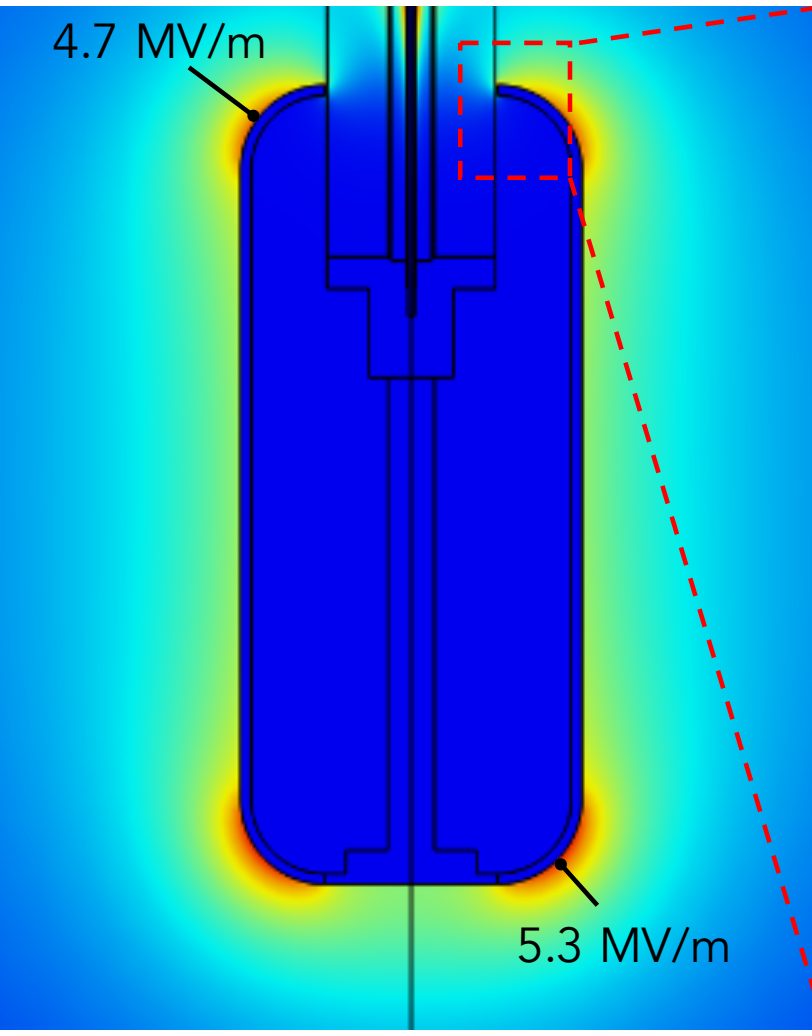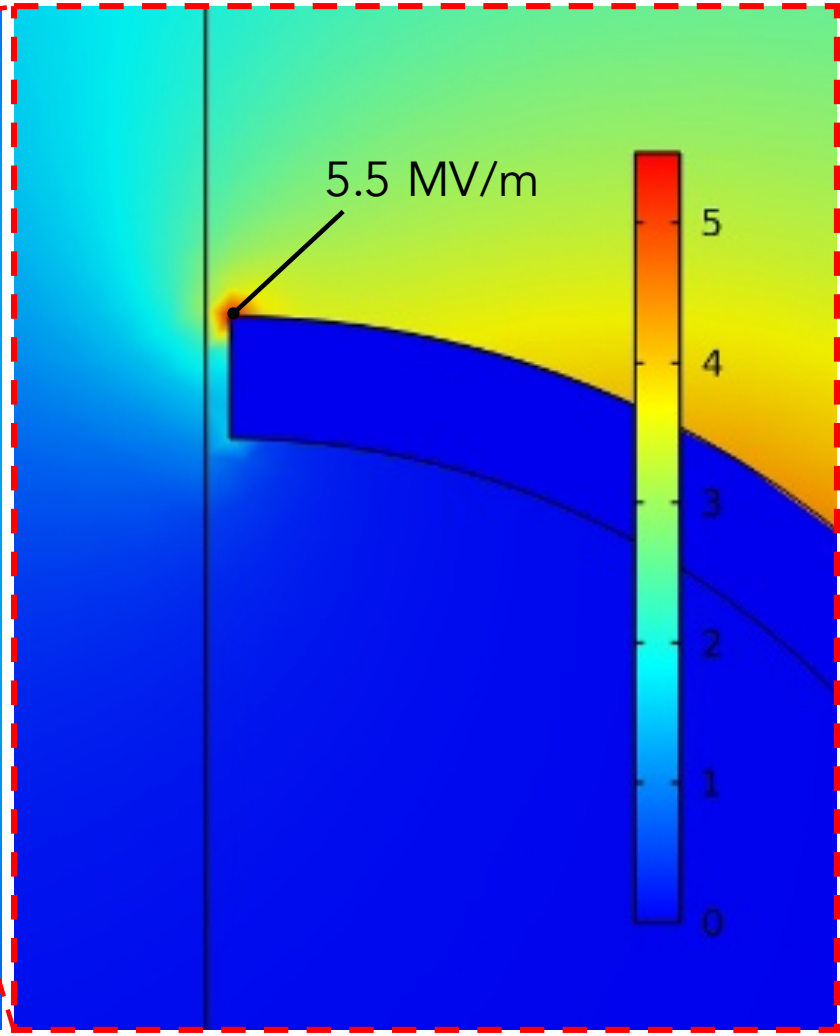

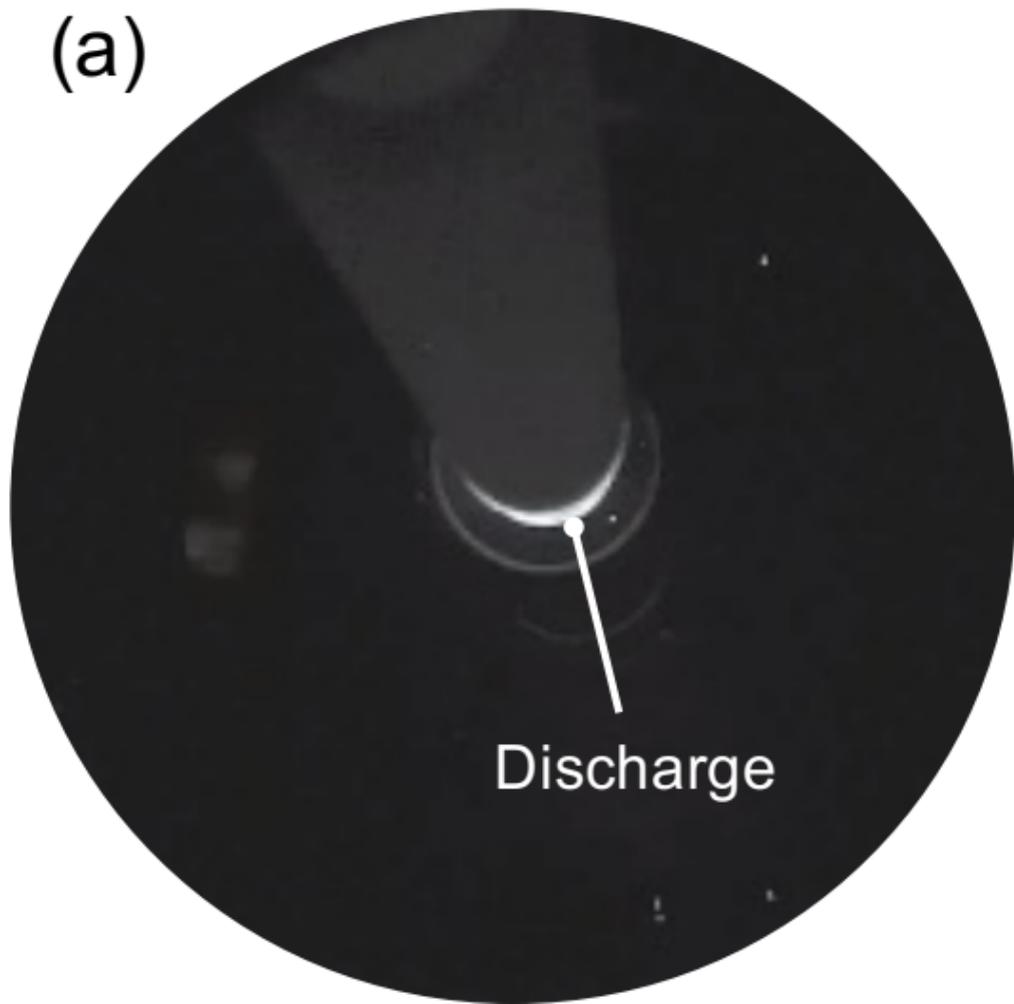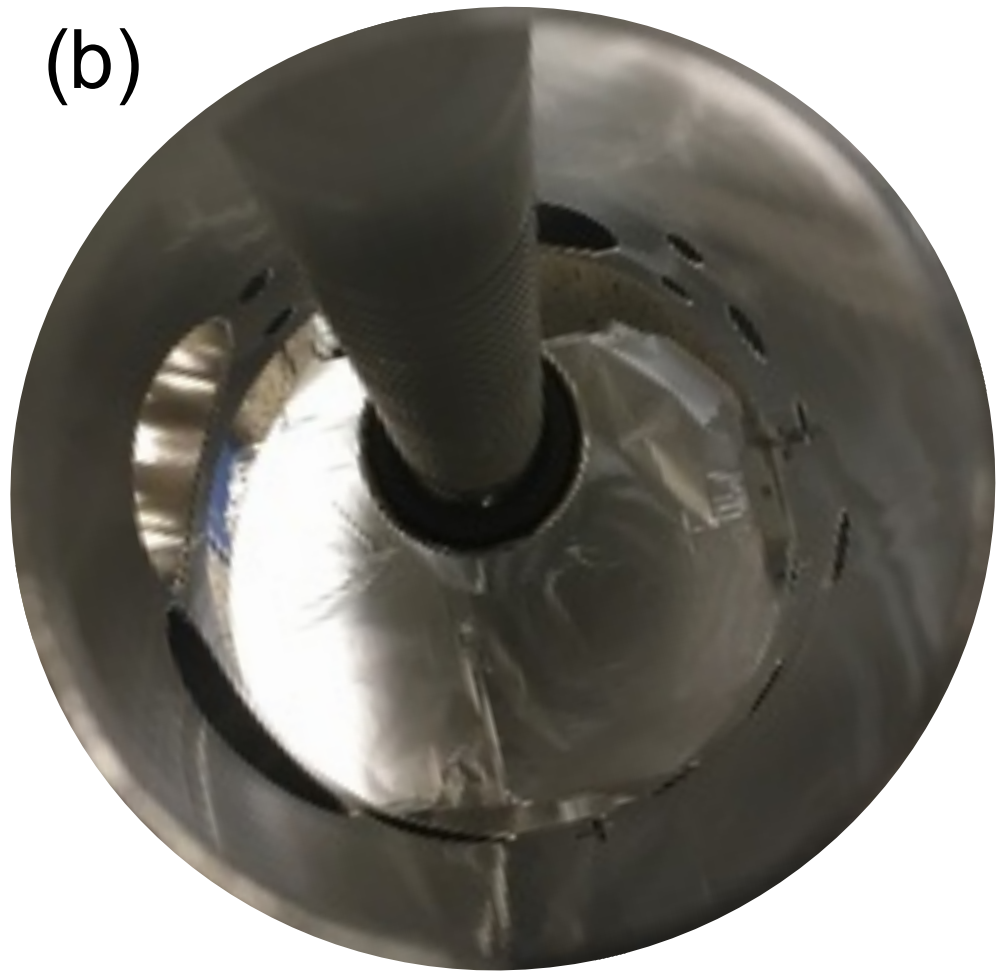